# BIG DATA ANALYTICS: THE STAKES FOR STUDENTS, SCIENTISTS & MANAGERS

*A management perspective*

K. Viswanathan Iyer
Dept. of Computer Sci. & Eng.
National Institute of Technology
Tiruchirapalli – 620015
e-mail: kvi@nitt.edu

**Abstract**

For a developing nation, deploying big data (BD) technology and introducing data science in higher education is a challenge. A pessimistic scenario is: Mis-use of data in many possible ways, waste of trained manpower, poor BD certifications from institutes, under-utilization of resources, disgruntled management staff, unhealthy competition in the market, poor integration with existing technical infrastructures. Also, the questions in the minds of students, scientists, engineers, teachers and managers deserve wider attention. Besides the stated perceptions and analyses perhaps ignoring socio-political and scientific temperaments in developing nations, the following questions arise: *How did the BD phenomenon naturally occur, post technological developments in Computer and Communications Technology and how did different experts react to it? Are academicians elsewhere agreeing on the fact that BD is a new science? Granted that big data science is a new science what are its foundations as compared to conventional topics in Physics, Chemistry or Biology? Or, is it similar in an esoteric sense to astronomy or nuclear science? What are the technological and engineering implications locally and globally and how these can be advantageously used to augment business intelligence, for example? In other words, will the industry adopt the changes due to tactical advantages? How can BD success stories be faithfully carried over elsewhere? How will BD affect the Computer Science and other curricula? How will BD benefit different segments of our society on a large scale?* To answer these, an appreciation of the BD as a science and as a technology is necessary. This paper presents a quick BD overview, relying on the contemporary literature; it addresses: characterizations of BD and the BD people, the background required for the students and teachers to join the BD bandwagon, the management challenges in embracing BD so that the bottomline is clear.

## 1. Introduction

The new Big-Data-(BD)-analytics-wave is currently perceived as a scientific Tsunami connoting different meanings and consequences to various stakeholders such as students, scientists, engineers, teachers, managers and policy makers at large. Some of the in-context questions still pertinent for many different segments of our society are summarized in Table 1 – in each case, an associated question asks: *Will I/we get assured returns on the investments?* The scenario from many advanced countries probably answers or have answered these questions in favor of the challenges in handling, analyzing and understanding huge scales of heterogeneous unstructured data originating from many different sources; in some places there are varying, and often fuzzy answers, however. A 2011 report on BD authored by McKinsey Global Institute, an economic research arm of McKinsey and Company highlighted BD analytics as a key driver in the next wave of economic innovation. The report suggests that the BD innovation may be slowed-down by a non-availability of skilled manpower essential to reap the insights from BD in various contexts. In the Indian scientific and business communities, the BD buzzword is slowly making way in many quarters. At a gross level, the BD industry that promises to throw-open new job opportunities, has four arms: **(a) Products wing (b) In-house analytics wing (c) Services and consultancy wing (d) Research and development wings.** Then, an instructive exercise for policy-makers, say for financial apportionment and planning, is to collect group-wise (e.g., state-wise, income-based, occupation-based) statistics based on Table 1. An industry estimate is that the BD analytics sector in India is a USD 2.00 billion industry - not counting the analytics training market - and it is expected to witness eight-fold growth between between now and 2020-22. Another estimated figure is that India holds 35% - 50% of global analytics services market. This kind of a market survey may be the deciding factor to make an alignment towards BD in a country like India. It may be argued by many that to get a broader perspective, a little more elaboration as regards the prevailing international scenario is required for policy-making minds. A clear regional understanding in terms of the following questions arising in different population segments (see relevant groups in Table 1) is essential, apart from stated perceptions, assumptions and analyses due to experts that may not be critically aware of the socio-political and scientific temperaments in developing nations:

> *How did the BD phenomenon naturally occur, post technological developments in Computer and Communications Technology and how did different experts react to it? Are academicians elsewhere agreeing on the fact that BD is a new science? Granted that big data science is a new science what are its foundations as compared to conventional topics in Physics, Chemistry or Biology? Or, is it similar in an esoteric sense to astronomy or nuclear science? What are the technological and engineering implications locally and globally and how these can be advantageously used to augment business intelligence, for example? In other words, will the industry adopt the changes due to tactical advantages? How can BD success stories be faithfully carried over elsewhere? How will BD affect the Computer Science and other curricula? How will BD benefit different segments of our society on a large scale?*

TABLE 1: Big data concerns among various polulation groups

| Pertinent questions and concerns: | Relevant groups: |
|---|---|
| Should we do further courses in probability, statistics, randomized algorithms, linear algebra, machine learning, data mining, data visualization techniques, software engineering, data-driven problem solving etc.? | Students and teachers of mainly Computer Science and Mathematics, Technical staff members in organizations. |
| Should we invest more time and money in learning and acquiring skills in one or more of *Python, SQL, Hadoop and MapReduce, Apache Spark, Hive, Pig, R, SPSS, Google chart, Microsoft Power BI, Tableau, D3 (data driven document), Fusion chart* etc. ? | Students and teachers of Computer Science, Some technical staff members in organizations. |
| Should we try out BD courses at undergraduate or postgraduate level or should we plan a 2-year or 4-year program under the BD banner or should we have a 4-year program wherein in the final year prescribed BD course-package will be offered as a specialization assuming that provision for the prerequisites are enabled earlier? | Heads of academic institutions, Some university professors as well as professors elsewhere. |
| Should we propel an international conference and workshop series on BD analytics? How many **MOOC**s are BD-related courses? | Heads of departments, Professors, Business houses. |
| How many certain high-valued ($-valued, for example) BD projects can we net in the next two to three years? | Middle-level managers and technical staff members in organizations and in some universities. |
| Should we create an eco-system for BD analysis by procuring new hardware platforms and architecture? | Top-level managers, Computer centre managers in universities and in commercial firms. |
| Should we have a national policy on big data analytics or should we react with high-power BD consortiums? Should a council for BD, ethics and society be publicised and empowered at a global level? | Think-tank brains, beaurocrats, administrators at the national and international levels. |

For a developing nation like India, an unanalysed deployment of any new technology under resource constraints and with a poor insight is bound to bring-in more disadvantages e.g., **Mis-use of data in many possible ways, waste of trained and qualified manpower, low-quality BD certifications from institutes, poor utilization of cost-intensive resources, disgruntled pool of management staff, unhealthy competition in the market, poor integration with existing technical infrastructure and associated scenarios, unethical uses of the new science.** This will lead to a more chaotic IT culture and society and the stated benefits of BD may not reach the deserving groups of our polulation on time.

This paper presents a comprehensive and quick BD overview as well as implications, relying on the contemporary literature; in particular the following points are addressed: popular characterizations and features of BD and the BD people, the skills and background required for the students as well as teachers to join the BD bandwagon, the possible management challenges in embracing BD to attain the promised gains with a clear bottomline. For a detailed overview of the state-of-the-art in BD in the last four years, reference is made to [7] which analyzes and classifies 457 papers concerning BD giving information to practioners and researchers on the trends in research and applications of BD in different domains with a reference overview of BD tools.

## 2. Understanding BD and their analysts and scientists

Among other sources, Internet and the social media are currently contributing to an unprecedented data and information explosion; the immediate consequence is summarized by John Naisbett:

> We are drowning in information and starving for knowledge.

A similar consequence is observed in [3] - the vast amount of data stored in some databases leads to the following phenomenon:

> We cannot see the wood viz., the patterns, for the trees viz., the individual data records.

Digitalization in almost all fields has resulted in many data sources and has contributed to extreme volumes of data some of which are of importance to gain meaningful knowledge from a statistical point of view at least in part. As an example, BD in the healthcare sector consists of the following data sources - electronic data plus other health records which may be in a protected-mode to maintain confidentiality of patients:

**Diagnostic data:** Images, ECG & EEG signals, blood, sputum and other test results.
**Instrumentation data:** RFID, barcodes, video feeds, sensors, monitors.
**Structured data:** ERP related, transactional data, clinical information systems, prescriptions, payment details.
**Unstructured data:** Consultation recordings, notes, patient instructions, social media discussions, diary notings.

The resulting data over some time period if covered-well and interpreted by domain experts, is of significance to the medical community to gain knowledge as well as to take decisions e.g., reducing readmission cases based on past-history. As a second example, researchers in human-computer interactions are known to collect rich behavioral traces of user interactions recorded as logs with online systems *in situ* on a BD scale. Both these examples point to new challenges in data collection and interpretation and ethics (see for example Table 1 in [6] to get an overview of the differences between medical big data analysis and classical statistical analysis). An attempt to characterize BD due to D. Laney in 2001 was in terms of the now-popular three V's viz., Volume, Variety, Velocity described below (see also [5]):

- *Volume* is taken to refer to the exponential growth in the overall size of the data relevant and accessible to an enterprise. The data collected is massive-enough to warrant new storage and access techniques as well as new algorithms for processing and analyzing the data corpus which may be distributed. One quoted figure capturing high volume is 60 GB of data creation in one second.
- *Velocity* refers to the speed at which data may be received as well as processed. Data processing now can mean batch processing, which works on historical data or stream processing, which analyzes data in real-time as it is generated. This also refers to the rate of change of data which is especially relevant in the context of stream processing.
- *Variety* refers to the multifarious and often incompatible, non-compliant (structured as well as unstructured) data formats. Incoming data may join a data corpus while coming from many different sources and in many different formats. Pre-processing is a requirement before analysis and that takes a significant amount of effort as well as time.

Subsequently, other V-attributes (e.g., *Veracity* - meaning uncertainty or inaccuracy present in the collected data repository) have been added in many similar situations. These are the attributes that are attributed to distinguish BD in BD analytics from the more well-known *science of analytics* used synonymously with *business intelligence* (traditional business intelligence tools work with structured data and they cannot handle the complexities of constantly changing data sources) in corporate environments. An alternative viewpoint from the data mining angle is presented as the **HACE**-theorem in [9] by capturing the characteristic features of the BD phenomenon. Thus data from the Internet, text data from facebook feeds, time and location data from GPS and mobile phones, smart grid and sensor data are all of interest to different customers in a BD scenario.

The use of the phrase *data science* was ushered at the beginning of the year 2000. The initial reaction within the scientific community was to question how and why it is different from Statistics and Information Science. Many scientists soon recognized that data-driven thinking, practice, education and research goes far beyond all previous efforts related to data processing of mainly well-structured data. To underscore the international importance of this new field one can note that the computational intelligence society of IEEE, U.S.A. has a technical committee on data mining and BD analytics - the overall goals of this technical committee are stated as

(a) *to promote the research, development, education and understanding the principles and applications of Data Mining and BD analytics and*

(b) *to help researchers whose background is primarily in Computational Intelligence in increasing their contributions to this area.*

Data analysts translate numbers into English, for instance – every business house collects data e.g., sales figures, market research, logistics, transportation costs. A data analyst takes company data and processess it so that the company is in a better position to take decisions – how a stock or a portfolio will perform given international financial market parameters, how to price new materials for the market, how to reduce transportation costs, how many people should work on a Saturday, how to distribute profits to enhance productivity. In the e-commerce business these business values of BD are recognized:

personalization, dynamic pricing, customer service, supply chain visibility, security and fraud detection, predictive analysis. Success stories have been reported by firms such as UPS, P&G, Amazon, PayPal (see [1]). The October 2012 edition of Harvard Busines Review article quoted what is attributed to both W. Edwards Deming and Peter Drucker: You cannot manage what you do not measure. The article goes on to say that because the BD managers can measure and hence know radically more about their businesses, they can directly translate their knowledge into better decision-making and performance. Descriptive, predictive and prescriptive analytics are known in business environments. This is in contrast with what has been called *HiPPO-based decisions* i.e., decisions due to highest-paid person's opinion. The following quote suggests the need for a working knowledge of statistics and the use of software tools by data scientists in busines environments:

> A data scientist is one who knows more statistics than any software engineer
> and one who knows more software engineering than any statistician.

Data analytics is the process of examining large data sets in order to draw conclusions about the information they contain, with the aid of specialised systems and software. Again, on the commercial side, modern data analytics initiatives can help businesses to increase revenues, improve operational efficiency, optimise marketing campaigns and customer service efforts, respond more quickly to market trends and gain a competative edge over rivals – all with the ultimate goal of boosting-up business performance and industrial process automation. In the Western scenario, the major challenges in making analytics work in the context of BD according to informal surveys include the following:

(a) Determining how to get the bang off BD
(b) Issues related to security, privacy, data quality
(c) Obtaining the needed skills and capabilities
(d) Integrating multiple data sources
(e) Integrating big data technology with existing IT infrastructure
(f) Funding needed for BD transformation from Pre-BD era to Post-BD era.

In many corporate environments there are Operations Analysts, Marketing Analysts, Financial Analysts using techniques in analytics (IBM Analytics is one example). For all these jobs, the foundational requirements are – Mathematics, Statistics and Computer Science; knowledge of economics and operations research and a degree in business management may be required in some cases. The following distinction is often made:

1. Data analysts are trainees or a juniors in the game. They use computer-based available tools and techniques.
2. Data scientists and Big Data engineers are experienced seniors who can create data and techniques for others.

The term data scientist is often attributed to J J Patil of LinkedIn analytics and J. Hammerbacher of Facebook analytics (Oct. 2012 edn., Harvard Business Review). We note that new knowledge is generated due to research in experiments, theoretical insights and due to computer simulations; Gordon Bell, Tony Hey and Alex Szalay have proposed that data-intensive science as a fourth dimension that contributes to new knowledge [2].

The role and the works of the analytics community in the post-BD era has prompted the following question on a larger scale:

> Are Ph.D.s in statistics useful? Are we ready for a doctorate in big data analytics?

On the hype-filled scientific front for top-class students, BD analytics are reported to assist researchers to verify or disprove scientific models, theories and hypotheses. It has been remarked that data science can be used to discover correlations (e.g., what phenomena occured) but cannot be used to establish causality (why the phenomena occured). As reported in the literature, BD analytics research employs *machine learning* (a paradigm developed by AI folks in the 1950s – the IBM researcher Arthur Lee Samuels apparently coined the term), data mining, statistics and visualization techniques to collect, process, analyze, visualize and interpret results to gain hidden knowledge in specific domains. Example domains of BD applications quoted in the literature include Computer Science, Engineering, Mathematics, Physics (e.g., 'Onics) and Astronomy, Biochemistry (e.g., 'Omics, genetics), Materials Science, Medicine. However in critical application domains the limits of insights from BD techniques cannot be ignored. To quote M. Hilbert, U. California, Davis (Jan. 2015), *"To predict a future that has never been, theory-driven models are necessary. . . . data mining and machine learning methods do not aim at providing such theories, they simply predict".* The fact is that a simple model may predict better than a sophisticated one (see for example, the linear regression example in [8]).

## 3. The BD thrust from students' and managers' viewpoints

### 3.1 The students' and teachers' view

As for a Computer Science job the usual business skills viz., analytical problem-solving, effective oral and writen communication and industry knowledge is relevant to a BD job too. This apart, data analysts with others assist is the following phases in order in the BD value-chain: data acquisition, data analysis, data curation, data storage. In the final phase, data analysts are expected to make realize the following usages of data: decision-support, prediction, simulation, exploration, visualisation, modeling, control, domain-specific usage. For an aspiring student of a typical BD job the high-level job profile can be summarized as under:

1. Work with IT teams, managements and/or data scientists to assist decision-making.
2. Mine data from primary and secondary sources.
3  Clean and prune data to discard irrelevant information.
4. Analyse and interpret results using ststistical tools and techniques.
5. Pinpoint trends, corelations and patterns in complicated data sets.
6. Identify new opportunities for process/product development.
7. Provide concise data reports and clear data visualizations for management.
8. Design, create, interface and maintain databases/data systems.

As a consequence of the BD job opportunities many teaching efforts are now being directed towards building competance for BD jobs. The University of Warwick course on *Foundations of Data Analytics* taught by the Computer Science department is an example from a non-US place - the Jan 2018 course-content highlights is given below:

- Introduction to analytics and contemporary case studies
- Basic tools for data manipulation and visualization
- Statistics with R
- Databases – Relational and NoSQL systems
- Regression – for linear case and for higher dimensional data
- Matrix analysis including Singular Value Decomposition and Principal Components Analysis
- Approaches to Clustering
- Classification models, including Support Vector Machine.
- Data Structures for complex data analysis
- Data Sharing, including k-anonymity, and differential privacy
- Graphs and its applications to social network data.

BD training schools focus on developing skills apart from imparting basic knowledge. Besides an earned basic degree, certifications can further help a prospective employee to demonstrate his or her expertise in the use of one or more analytical tools used for BD work. Typical data science training courses are now organized around the R programming language with exposure to visualization using graphics.

The ongoing initiative at the National Institute of technology, Tiruchirapalli (NIT, Trichy) is a case that may be relevant to other new programs for BD in similar institutions. At the joint secretary level, the Ministry of Human Resource Development, Government of India initiated in Dec. 2016 the need for all NITs to launch new programs and courses in emerging priority areas including Computer Science and Information Technology, an area that is still evolving, flourishing and changing the global landscape. A BD analytics course at the Master's level under the engineering stream, in the pipeline, was started in 2017 at NIT, Trichy. Table 2 given below, prepared at that time but still relevant, summarizes the author's concerns in announcing any new BD program.

### 3.2 The challenges for managers

The transition to BD technology is an uphill task for rmanagement teams in many places. While legacy technologies could be replaced with smarter equivalents in many business domains, there are numerous reasons why substitution may not be a viable option in some contexts. A list of these impediments are provided below with respect to the manufacturing sector (see also [4]) – these may be relevant for others too:

1. Over 40 years of investment in legacy IT infrastructure and automation networks.
2. Regulatory policies and quality standards for new technology as in the case of medical industry.
3. Practice of in-place proprietary systems or protocols as opposed to open standards.
4. Weak vision and commitment i.e., a non-progressive leadership.
5. High risk and disruption associated with the adoption of new technology and system implementation.
6. Weakness in required skills and poor technology awareness.
7. Absence of a strong multi-disciplinary workforce.

A hard task for large industrial management teams is to squarely address the prevailing *digital divide*. The BD digital divide is: (d1) who creates the data (d2) who accesses the data and (d3) who has the resources to analyse the data (M. Hilbert 2013) – depending on different domain of applications, data privacy and information sharing mechanisms between data producers and data consumers can be significantly different (see [9]). Hence the need to have organizational mechanisms and appropriate management policies to harness the power of BD is a requirement for any success story. In the BD era, the major management challenges as identified by Andrew McAfee and Erik Brynjolfsson both of M.I.T. Sloan School of Management, U.S.A., are the following:

- Good and strong business leadership that can turn perceived opportunities into real values for the stakeholders.
- Effective decision making by putting information and the relevant decision rights in the same location.
- Providing the right pre-requisites, tools and techniques e.g., statistical knowledge, visualization tools for the BD challenges.
- Making available the relevant technology e.g., the Hadoop framework using commodity hardware and open-source software.

The final point of concern for any management should be *company culture* -- to quote McAfee and Brynjolfsson, "The first question a data-driven organization asks itself is not *What do we think?* but *What do we know?* This requires a move away from acting solely on hunches and instinct. It also requires breaking a bad habit we've noticed in many organizations: pretending to be more data-driven than they actually are. Too often, it is the case that executives who spiced up their reports with lots of data that supported decisions they had already made using the traditional HiPPO approach.  A desirable company culture will promote performance over compliance, with an emphasis on minimising energy and material usage, while maximising sustainability,  health and safety, and economic competitiveness. Transforming operations from reactive and  responsive, to those that are predictive and preventative is another profitable practice for oragnizations". Thus preparedness for participation in the BD game in developing nations should necessarily take into consideration, the prevailing IT climate i.e., the perceived effects of legacy and the associated people who contribute to it and who have control over it. For what the BD revolution is getting initiated and who all are getting benefitted is a relevant question for administrators. For any developing country like India, economic growth should not be skewed; it should lead to better labor market conditions, given that the majority of the workers in India are in informal jobs[1] (see India Labor Market Update, International Labour Organization, July 2016; for more refer to www.ilo.org/newdelhi). Agriculture, construction, manufacturing, medical industry and  global trade and exports are undoubtedly priority areas in India. It remains to be seen as to what extent BD can influence and eventually contribute to these areas, directly or indirectly.

TABLE 2. **Planning-stage concerns for a DB program launch**

| CONCERN | RELEVANT POINTS/QUESTIONS FOR ELABORATION |
|---|---|
| Types of inputs necessary to plan the program and get it started | Computer Science related background as the eligibility to join the program:<br>Minimum set of pre-requisite courses/equivalents and grades or marks to be accepted in transcripts, admission test score criteria if applicable and statutary reservation details.<br>Student strength related: Number of seats in a class and available hostel/dormitary accommodation, available scholarships and their sources for the new comers.<br>Resources: Classroom and laboratory spaces and schedules, new hardware and software, prescribed texts (electronic and cheap editions) and funding for these, journals for the library (online or print). |
| Types of output scenarios envisaged | What are the valid or validated employment and other opportunities here and abroad in web technology, analytics, BD?<br>Are there details of recruiting organizations such as company turnover, nature of clients, type of operations, employe statistics, salary ranges, whether involved in classified work?<br>By what mechanisms others will treat the given degree certificates on par with traditional degrees in Computer Science for employment or for higher studies?<br>What are the quality check parameters to sustain the program? |
| Faculty strength | Has any staff member completed any compact workshop or Coursera courses in any leading institute related to BD?<br>What are the chief areas of interests, including reaearch, of the teachers as seen from their publications?<br>Have the faculty members offered isolated elective courses earlier and if so are the lecture notes available?<br>Are there consent letters from visiting teachers or industry experts who can offer package courses? Are there tie-ups with others?<br>To start a new program are new hands identified? What are the current teaching and administrative loads for the teachers?<br>Are there plans to drop one or more existing program?<br>Are computer laboratory assistants and programmers geared-up? |
| Note on similar programs elsewhere | How will this program compare and compete with other programs in Computer Science or Applications?<br>Are there any statistics study with confidence to inform how many of the student population (with adequate background in Computer Science, Computer Applications, Electronics, Mathematics) will opt for this course? |
| Curriculum and Syllabus | Apart from the following core courses suggestive for the BD program sought, what other relevant mandatory courses are planned?<br><br>    Calculus, Discrete Mathematics, Numerical Computing, Linear Algebra,<br>    Probability Theory, Statistics.<br>    Python Programming, R Programming, Data Structures, Algorithms,<br>    Databases Systems, Randomized Algorithms, Web Technology and XML.<br>    Data Mining, Data Science, Machine Learning, Hadoop/MapReduce or<br>    Apache Spark.<br>    Parallel Computing, Distributed Computing, Distributed Databases.<br><br>Are the prerequisites and other courses partially ordered and are credit-hours for theory and practice identified?<br>Are say 30-hours lecture samples available for the new courses?<br>Are typical case studies well-understood with respect to say, financial analysis, market survey, scientific computing (e.g., clustering web documents, data cleaning) etc.?<br>Which are the courses that may be credited by outside-program students? |

**Acknowledgement.** The author would like to thank the anonymous reviewer for the suggestions on the earlier draft.

---

1   The National Commission for Enterprises in Unorganized Sector in India gives this definition (2008): all unincorporated private enterprises owned by individuals or households engaged in the sale and production of goods and services operated on a proprietary or partnership basis with less than ten workers. The informal workers are defined thus: those working in the unorganized sector or households, excluding regular workers with social security benefit provided by the employers and the workers in the formal sector without any employment and social security benefits provided by the employers.